\documentstyle[12pt,epsf]{article}
\newcommand{\nn}{\nonumber}
\newcommand{\be}{\begin{equation}}
\newcommand{\ee}{\end{equation}}
\newcommand{\bea}{\begin{eqnarray}}
\newcommand{\eea}{\end{eqnarray}}

\def\bfnabla{\mbox{\boldmath $\nabla$}}

\def\als{\alpha_{\rm s}}
\def\siml{{\ \lower-1.2pt\vbox{\hbox{\rlap{$<$}\lower6pt\vbox{\hbox{$\sim$}}}}\ }} 

\def\lla{\langle\!\langle}
\def\rra{\rangle\!\rangle}

\newcommand{\Appendix}[1]%
    {%
     \section{#1}%
      }

\newcommand{\AmS}{{\protect\the\textfont2 A\kern-.1667em\lower.5ex\hbox{M}\kern-.125emS}}

\begin{document}
\begin{titlepage}
\begin{flushright}
\tt
HD-THEP-01-14 \\
IFUM-678-FT
\end{flushright}

\vspace{1cm}
\begin{center}
\begin{Large}
{\bf Poincar\'e invariance and the heavy-quark potential}
\\[2cm]
\end{Large} 
{\large Nora Brambilla\footnote{nora.brambilla@cern.ch}}\\
{\it Dipartimento di Fisica, Universit\`a degli Studi di Milano\\
     via Celoria 16, 20133 Milano, Italy} \\[1cm]
{\large Dieter Gromes\footnote{d.gromes@thphys.uni-heidelberg.de}}\\
{\it Institut f\"ur Theoretische Physik, Universit\"at Heidelberg\\
     Philosophenweg 16, 69120 Heidelberg, Germany}\\[1cm]
{\large Antonio Vairo\footnote{antonio.vairo@cern.ch}}\\
{\it Theory Division CERN, 1211 Geneva 23, Switzerland}\\[1cm]
\end{center}

\vspace{1cm}
\begin{abstract}
\baselineskip=16 pt 
{We derive and discuss the constraints induced by Poincar\'e invariance on the form of the 
heavy-quark potential up to order $1/m^2$. We present two derivations: one uses general arguments 
directly based on the Poincar\'e algebra and the other follows from an explicit calculation 
on the expression of the potential in terms of Wilson loops.
We confirm relations from the literature, but also clarify the
origin of a long-standing false statement pointed out recently.
\vspace{0.5cm}\\ }
\end{abstract}

\end{titlepage}

\vfill\eject

\setcounter{footnote}{0}
\pagenumbering{arabic}

\section{Introduction}
The effective field theory (EFT) approach has recently clarified under which circumstances 
the heavy quark-antiquark interaction may be described by a simple potential picture. 
We call pNRQCD (potential non-relativistic QCD) the corresponding EFT. Moreover, in this framework 
the {\it complete} $1/m^2$ quark-antiquark potential has been derived \cite{m1,m2}. 
The final expression of the potential is given in terms of Wilson loops \cite{wilson}.
Therefore, it is appropriate also in the situation where the quark-antiquark 
interaction is non-perturbative. In this case the Wilson loop operators 
may be calculated on the lattice \cite{bali} or in QCD vacuum models \cite{bv}.

With respect to the classical literature on the subject (for a review see \cite{rev})  
the derivation \cite{m1,m2} presents the following improvements.
{\it i)} Being done in an EFT framework it correctly implements the 
ultraviolet behaviour of QCD. This is encoded into the matching coefficients 
inherited from the matching to NRQCD \cite{NRQCD} (on the relation between the matching coefficients 
of NRQCD and the heavy-quark potential see also \cite{chen}). 
{\it ii)} Besides the spin and momentum-dependent potentials,  
it includes the $1/m$ and $1/m^2$ momentum and spin-independent potentials that have not 
been calculated before \cite{ef,bmp}. Hence, the obtained expression is complete. {\it iii)} 
It corrects some errors present  in the previous literature, which have propagated in several 
related papers until today (we refer to \cite{m2} for a discussion on this point).

In view of these new information (and in some case corrections) with respect to the previous 
knowledge on the subject, one has to ask if other results, present in 
the literature and related to the form of the heavy-quark potential, may change.   
In particular in \cite{poG} and \cite{poBBP} exact relations among the spin and 
momentum-dependent potentials, respectively, have been derived.
A more recent analysis can be found in \cite{pochen}. One may wonder if the 
errors found in the previous literature and/or the new terms added to the potential 
spoil the validity of those exact relations. In the present paper we will address 
this question and derive {\it ex novo}, by general considerations on the structure 
of the potential and by explicit manipulations of the expressions given in \cite{m1,m2}, 
the relations induced by Poincar\'e invariance on the form of the potentials.
We shall confirm the results found in \cite{poG,poBBP} and explain why the errors found 
in the literature on the derivation of the potentials do not affect the derivation 
of the exact relations among them. Since the potentials are defined up to unitary 
transformations, we shall also show that the relations found are, indeed, invariant with 
respect to them.

The consequences of the above analysis, i.e. the obtained exact relations among the 
heavy-quark potentials, may be relevant in several situations. For instance,  
they allow to have independent checks on analytic, and lattice computations. 
However, in our opinion, the most important application is to establish and 
constrain the power counting of NRQCD under the specific situations 
where this EFT may be substituted by pNRQCD. This is an important point not only 
in the study of the quarkonium spectrum, for instance by means of lattice NRQCD, 
but also in the study of the quarkonium production where the validity 
of the perturbative NRQCD power counting has been recently questioned \cite{m2,roth}.

The paper is organized as follows. In Sec. \ref{secpnrqcd} we shortly describe 
the physical situation we are considering and write down the corresponding pNRQCD Lagrangian. 
In Sec. \ref{secpo} we discuss Poincar\'e invariance from a general point of view and derive 
relations among the different potentials without specifying their explicit form. 
In Sec. \ref{secex} the previous relations are proven to be valid by direct computation on the potentials 
explicitly given in \cite{m1,m2}. In Sec. \ref{seccon} we discuss the obtained results. 
In an appendix we clarify in some detail the origin of the error in the previous literature, 
which was found in \cite{m2}, and its relation with the derivation of Sec. \ref{secex}.

\section{pNRQCD}
\label{secpnrqcd}
In this work we consider pNRQCD in the situation where the singlet field $S$, describing the 
heavy-quarkonium system, is the only available ultrasoft degree of freedom. 
In the perturbative regime this situation corresponds to considering pNRQCD at the leading order in the 
multipole expansion while in the non-perturbative one it corresponds to considering pNRQCD without 
light quarks and ultrasoft gluonic excitations under the circumstances discussed  in \cite{m1,m2}. 
The pNRQCD Lagrangian reads
\be
{\cal L}_{\rm pNRQCD} = S^\dagger 
\bigg( i\partial_0 -h_s({\bf x}_1,{\bf x}_2, {\bf p}_1, {\bf p}_2)\bigg) S, 
\label{pnrqcdl}
\ee
where ${\bf x}_j$ and ${\bf p}_j$ are the position and the momentum operators of a heavy quark 
of mass $m_j$ and spin ${\bf S}_j$. The operator $h_s$ can be identified with the Hamiltonian 
of the singlet and has the following structure\footnote{
This structure already implements some of the constraints, like translational invariance, 
coming from the following discussion on the Poincar\'e group.} up to order $1/m^2$
\bea
&& \hspace{-14mm}
h_s({\bf x}_1,{\bf x}_2, {\bf p}_1, {\bf p}_2)
 ={{\bf p}^2_1\over 2 m_1} +{{\bf p}^2_2\over 2 m_2} + V^{(0)}(r)\nn\\
&& \hspace{18mm}
+{V^{(1,0)}(r) \over m_1}+{V^{(0,1)}(r) \over m_2}+ {V^{(2,0)} \over m_1^2}
+ {V^{(0,2)}\over m_2^2}+{V^{(1,1)} \over m_1m_2}, \label{hs}\\
&& \hspace{-14mm}
V^{(2,0)} = {1 \over 2}\left\{{\bf p}_1^2,V_{{\bf p}^2}^{(2,0)}(r)\right\}
+{V_{{\bf L}^2}^{(2,0)}(r)\over r^2}{\bf L}_1^2
+ V_r^{(2,0)}(r) + V^{(2,0)}_{LS}(r){\bf L}_1\cdot{\bf S}_1, \label{v20}\\
&& \hspace{-14mm}
V^{(0,2)} = {1 \over 2}\left\{{\bf p}_2^2,V_{{\bf p}^2}^{(0,2)}(r)\right\}
+{V_{{\bf L}^2}^{(0,2)}(r)\over r^2}{\bf L}_2^2
+ V_r^{(0,2)}(r) - V^{(0,2)}_{LS}(r){\bf L}_2\cdot{\bf S}_2, \label{v02}\\
&& \hspace{-14mm}
V^{(1,1)} = -{1 \over 2}\left\{{\bf p}_1\cdot {\bf p}_2,V_{{\bf p}^2}^{(1,1)}(r)\right\}
-{V_{{\bf L}^2}^{(1,1)}(r)\over 2r^2}({\bf L}_1\cdot{\bf L}_2
+ {\bf L}_2\cdot{\bf L}_1)+ V_r^{(1,1)}(r) \nn \\ 
&& \hspace{-14mm}
+ V_{L_1S_2}^{(1,1)}(r){\bf L}_1\cdot{\bf S}_2 - V_{L_2S_1}^{(1,1)}(r){\bf L}_2\cdot{\bf S}_1
+ V_{S^2}^{(1,1)}(r){\bf S}_1\cdot{\bf S}_2 + V_{S_{12}}^{(1,1)}(r){\bf S}_{12}({\hat {\bf r}})
\label{v11}, 
\eea
where ${\bf r} = {\bf x}_1-{\bf x}_2$, ${\bf L}_j \equiv {\bf r} \times {\bf p}_j$ and 
${\bf S}_{12}({\hat {\bf r}}) \equiv 12 {\hat {\bf r}}\cdot {\bf S}_1 \,{\hat {\bf r}}\cdot {\bf S}_2  
- 4 {\bf S}_1\cdot {\bf S}_2$.  The explicit form of the potentials has been determined 
in \cite{m1,m2} in terms of Wilson loop operators and will be the subject 
of Sec. \ref{secex}.

\section{Poincar\'e Invariance}
\label{secpo}
Given a specific system, the generators ${\bf P}$ of space translations, 
the generator $H$ of time translations, the generators ${\bf J}$  of rotations 
and the generators ${\bf K}$ of Lorentz transformations satisfy the well-known  
Poincar\'e algebra (some original references can be found in \cite{Foldy}) 
\begin{eqnarray}
[{\bf P}_i,{\bf P}_j] &=&0, \label{A1}\\ 
{[{\bf P}_i,H]} &=& 0, \label{A2}\\ 
{[{\bf J}_i,{\bf P}_j]} &=& i \epsilon^{ijk}{\bf P}_k , \label{A3}\\  
{[{\bf J}_i,H]} &=& 0, \label{A4}\\  
{[{\bf J}_i,{\bf J}_j]} &=& i \epsilon^{ijk}{\bf J}_k , \label{A5}\\  
{[{\bf P}_i,{\bf K}_j]} &=&-i \delta_{ij} H , \label{A6}\\  
{[H,{\bf K}_i]} &=&-i {\bf P}_i , \label{A7}\\  
{[{\bf J}_i,{\bf K}_j]} &=& i \epsilon^{ijk}{\bf K}_k , \label{A8}\\  
{[{\bf K}_i,{\bf K}_j]} &=&-i \epsilon^{ijk}{\bf J}_k . \label{A9}
\end{eqnarray}
Let us consider a two-particle system of the type presented in Sec. \ref{secpnrqcd}.
We may identify the generators ${\bf P}$  and ${\bf J}$ with the total momentum 
and the total angular momentum of the system respectively:
\begin{eqnarray}
{\bf P} &=& {\bf p}_1 + {\bf p}_2, \qquad {\bf p}_j = -i \bfnabla_{{\bf x}_j},  \label{P2}\\
{\bf J} &=& {\bf x}_1 \times {\bf p}_1 + {\bf x}_2 \times {\bf p}_2 + {\bf S}_1 + {\bf S}_2 . \label{J2}
\end{eqnarray}
These definitions fulfill Eq. (\ref{A1}), (\ref{A3}) and (\ref{A5}).  
The meaning of the conditions set by Eqs. (\ref{A2}) and (\ref{A4}) is trivial. The first 
equation constrains $H$ to be a function of ${\bf x}_1-{\bf x}_2$, i.e. of the distance ${\bf r}$ 
of the two particles, only. The latter constrains $H$ to be a scalar under rotations. 
Of the remaining four commutators only three are independent \cite{Foldy2}.\footnote{
For instance, by using the Jacobi identity, it can be proven that Eq. (\ref{A7}) follows  
from Eqs. (\ref{A6}), (\ref{A8}) and (\ref{A9}). However, we will not make explicit use of this fact 
in the following. The reason is that we shall analyze the Poincar\'e algebra in the framework 
of a $1/m$ expansion. In this framework, when using $H$ and ${\bf K}$ up to a fixed order in $1/m$, 
the commutation relations are not all verified at the same order in $1/m$ \cite{yun}.} 
The constraints that they put on the form of the generators is by far less trivial and 
will be the main subject of this section. 

We consider our two particle system to be described by the Lagrangian (\ref{pnrqcdl}) and we 
identify the generator $H$ with 
\begin{equation}
H ({\bf x}_1,{\bf x}_2, {\bf p}_1, {\bf p}_2) =
m_1 + m_2 + h_s({\bf x}_1,{\bf x}_2, {\bf p}_1, {\bf p}_2).  
\label{HH}
\end{equation}
It has been proved in \cite{Foldy} that ${\bf K}$ can be written as:
\bea
{\bf K} &=& {1\over2}\sum_{i=1}^2 \left[
\left\{{\bf x}_i, m_i + {{\bf p_i}^2\over 2 m_i} + \dots \right\} 
- {{\bf S}_i\times {\bf p}_i \over m_i}(1  + \dots)\right]  - t{\bf P}  + {\bf U},
\label{K2one}
\eea
where $\{A,B\} =  AB + BA$, and the dots indicate higher order terms in the inverse of the mass.   
These are given in closed form in the quoted literature, but do not matter here.\footnote{One may wonder 
about the origin of the term $- t{\bf P}$ in Eq. (\ref{K2one}). It is a consequence 
of the identification of Eq. (\ref{HH}) and the request of Lorentz invariance for the Lagrangian 
(\ref{pnrqcdl}) (or, which is the same, for the Schr\"odinger equation). 
From the commutation relation (\ref{A7}) it then follows that the $t$ dependence of ${\bf K}$ 
is carried by a term $-t {\bf P}$.} ${\bf U}$ is a function, which, due to Eq. (\ref{A8}), has to be 
a vector under rotations. Moreover, it is related to $h_s$ via Eq. (\ref{A6}). 
Up to order $1/m$ a function ${\bf U}$ that satisfies the above relations is   
\bea
{\bf U} &=& 
{\bf x}_1 \left( v_a^{(0)}(r) + {v_a^{(1,0)}(r)\over m_1} + {v_a^{(0,1)}(r)\over m_2} \right) 
+ {\bf x}_2 \left( v_b^{(0)}(r) + {v_b^{(1,0)}(r)\over m_1} + {v_b^{(0,1)}(r)\over m_2} \right) \nn\\
& & + {\bf U}^{(0)} + {{\bf U}^{(1,0)}\over m_1} + {{\bf U}^{(0,1)}\over m_2} + \dots, 
\label{UU}
\eea
where the functions $v_{a,b}$ have to fulfill the conditions
\bea
V^{(0)}(r) &=& v_a^{(0)}(r) + v_b^{(0)}(r), \nn \\
V^{(1,0)}(r) &=& v_a^{(1,0)}(r) + v_b^{(1,0)}(r), \label{UUbis} \\
V^{(0,1)}(r) &=& v_a^{(0,1)}(r) + v_b^{(0,1)}(r). \nn
\eea
The functions ${\bf U}^{(0)}$, ${\bf U}^{(1,0)}$ and  ${\bf U}^{(0,1)}$ 
are arbitrary, but ${\bf K}$ obtained from them has to fulfill 
Eqs. (\ref{A6})-(\ref{A9}). In particular, they are vectors and commute with ${\bf P}$, i.e. 
they depend on momenta and the relative distance ${\bf r}$ only \cite{yun}. 
Moreover, ${\bf U}^{(0)}$ commutes with $V^{(0)}$.

The relevant point is the following. $H$ is now established up to order $1/m^2$, Eqs. (\ref{HH}) and 
(\ref{hs}), and ${\bf K}$ up to order $1/m$, Eqs. (\ref{K2one}) and (\ref{UU}). Then Eqs. (\ref{A6}), 
(\ref{A7}) and (\ref{A8}) have to be fulfilled up to order $1/m$ and Eq. (\ref{A9}) up 
to order $1/m^0$ (consider that ${\bf K}$ starts from order $m$).
The above definitions of $H$ and ${\bf K}$ trivially satisfy Eqs. (\ref{A6}), (\ref{A8}) and (\ref{A9}).
The situation is different for Eq. (\ref{A7}). In order that this relation holds at order 
$1/m$ some non-trivial relations between the potentials of Eq. (\ref{hs}) have to be fulfilled.
Therefore, as already pointed out in \cite{Dirac}, the Poincar\'e algebra may  
serve to constrain the form of the potentials. More specifically, from 
Eq. (\ref{A7}), taken up to order $1/m$, after a lengthy calculation, we obtain relations 
in the spin sector of the heavy-quark potential, 
\bea
& & \hspace{-5mm}
V_{LS}^{(2,0)}(r) - V_{L_2S_1}^{(1,1)}(r) + 
{1\over 2 r}{d V^{(0)}(r) \over dr} = 0, 
\label{spin1}\\
& & \hspace{-5mm}
V_{LS}^{(0,2)}(r) - V_{L_1S_2}^{(1,1)}(r) + {1\over 2 r}{d V^{(0)}(r) \over dr} = 0, 
\label{spin2}
\eea 
and in the spin-independent sector, 
\bea
& & \hspace{-5mm}
V_{{\bf L}^2}^{(2,0)}(r) + V_{{\bf L}^2}^{(0,2)}(r) 
- V_{{\bf L}^2}^{(1,1)}(r) + {r\over 2}{d V^{(0)}(r) \over dr} =0, 
\label{bmp1}\\
& & \hspace{-5mm}
-2(V_{{\bf p}^2}^{(2,0)}(r) + V_{{\bf p}^2}^{(0,2)}(r)) 
+ 2 V_{{\bf p}^2}^{(1,1)}(r) - V^{(0)}(r)+ r {d V^{(0)}(r) \over dr} =0, 
\label{bmp2}\\
& & \hspace{-5mm}
{1\over r} (V_{{\bf L}^2}^{(2,0)}(r) - V_{{\bf L}^2}^{(0,2)}(r)) 
+ {d\over dr}\big(V_{{\bf L}^2}^{(2,0)}(r) - V_{{\bf L}^2}^{(0,2)}(r)\big) 
+ {d\over dr}\big(V_{{\bf p}^2}^{(2,0)}(r) - V_{{\bf p}^2}^{(0,2)}(r)\big) = 0.
\label{bmp3}
\eea 
Notably the dependence on ${\bf U}^{(0)}$, ${\bf U}^{(1,0)}$ and  ${\bf U}^{(0,1)}$ 
disappears in the final relations (for more details see appendix \ref{appA}). If the system is invariant under 
$m_1 \leftrightarrow m_2$ transformation (and considering that one of the two particles 
is an antiparticle), we have the additional conditions 
\bea
V_{{\bf p}^2, \, {\bf L}^2}^{(2,0)}(r) &=& V_{{\bf p}^2, \,{\bf L}^2 }^{(0,2)}(r),  \label{sym1}\\
V_{r,\, LS}^{(2,0)}(r) &=& V_{r,\, LS}^{(0,2)}(r;m_2\leftrightarrow m_1), \label{sym2}\\ 
V_{L_1S_2}^{(1,1)}(r) &=& V_{L_2S_1}^{(1,1)}(r; m_1 \leftrightarrow m_2). \label{sym3}
\eea
In this specific situation Eqs. (\ref{spin1}) and (\ref{spin2}) reduce, up to the NRQCD 
matching coefficients, to the relation obtained in \cite{poG} and Eqs. ({\ref{bmp1})-(\ref{bmp3}) to the 
relations obtained in \cite{poBBP}. In both cases the proof was done by explicit calculation 
on the potentials expressed in terms of Wilson loops (see Sec. \ref{secex}). 
A proof derived from reparameterization invariance on the form of the scattering 
amplitude can be found in \cite{pochen}. 

We note that constraints on the potentials $V^{(1,0)}$, $V^{(0,1)}$, $V_r$, $V_{S^2}$ and $V_{S_{12}}$ 
could be obtained in this way only by including terms of order $1/m^3$ in the Hamiltonian (\ref{hs}). 
These terms are beyond our present knowledge.

\subsection{Unitary Transformations}
The Hamiltonian and hence the potentials are defined up to unitary transformations.
One may wonder if this ambiguity in the definition of the potentials  also 
affects the above exact relations. We will show, for the example of the 
unitary transformation considered in \cite{m1}, that this is not the case and 
that the relations derived from Poincar\'e invariance are invariant under unitary transformations.

In \cite{m1} the following unitary transformation has been considered:
\be
U = \exp\left( - i \left\{ {\bf W}({\bf r}), {{\bf p}_1\over m_1} - {{\bf p}_2\over m_2} \right\} \right),  
\label{unitary}
\ee
which transforms $h_s \to h_s^\prime = U^\dagger \, h_s \, U$.
We will check explicitly that the relations (\ref{spin1})-(\ref{bmp3}) 
are preserved by the transformation (\ref{unitary}). 
Up to order $1/m^2$, $h_s^\prime$ is given by:
\begin{eqnarray}
h_s^\prime &=& h_s 
- {1\over 2 m_1^2}\{{\bf p}^i_1,\{ {\bf p}^j_1, (\nabla^i_r W^j) \}\} 
+ {1\over2  m_1m_2}\{{\bf p}^j_1,\{ {\bf p}^i_2, (\nabla^i_r W^j) \}\} \nn\\ 
& & - {1\over 2 m_2^2}\{ {\bf p}^i_2,\{ {\bf p}^j_2, (\nabla^i_r W^j) \}\}
+ {1\over2  m_1m_2}\{{\bf p}^i_1,\{ {\bf p}^j_2, (\nabla^i_r W^j) \}\} \nn\\  
& & + \hbox{momentum independent terms}.
\label{hap2}
\end{eqnarray}
Let us now assume a decomposition of the form 
$\nabla^i_r W^j = {\hat r}^i{\hat r}^jA(r) + \delta^{ij}B(r)$. 
Then the momentum-dependent potentials of $h_s^\prime$ can be written  as 
\begin{eqnarray*}
V_{{\bf L}^2}^{(1,1)}(r)^\prime &=& V_{{\bf L}^2 }^{(1,1)}(r) + 4A(r), \\
V_{{\bf L}^2}^{(2,0)}(r)^\prime &=& V_{{\bf L}^2 }^{(2,0)}(r) + 2A(r), \\
V_{{\bf L}^2}^{(0,2)}(r)^\prime &=& V_{{\bf L}^2 }^{(0,2)}(r) + 2A(r), \\
V_{{\bf p}^2}^{(1,1)}(r)^\prime &=& V_{{\bf p}^2 }^{(1,1)}(r) - 4A(r) -4B(r),\\
V_{{\bf p}^2}^{(2,0)}(r)^\prime &=& V_{{\bf p}^2 }^{(2,0)}(r) - 2A(r) -2B(r),\\
V_{{\bf p}^2}^{(0,2)}(r)^\prime &=& V_{{\bf p}^2 }^{(0,2)}(r) - 2A(r) -2B(r).
\end{eqnarray*}
It then follows immediately that also the potentials obtained after the unitary transformation 
satisfy the relations (\ref{spin1})-(\ref{bmp3}).

\section{Relations obtained via Explicit Transformations of the Operators}
\label{secex}
In this section we will prove the relations (\ref{spin1})-(\ref{bmp2}) under 
the conditions (\ref{sym1})-(\ref{sym3}) by explicit transformation of the potentials,
\bea 
V^{(0)}(r) \!\!&=&\!\! \lim_{T\to\infty}{i\over T} \ln \langle W_\Box \rangle, \label{v0}\\
V_{LS}^{(2,0)}(r) \!\!&=&\!\!  {c_F^{(1)} \over 2r^2}i {\bf r}\cdot \lim_{T\rightarrow \infty}
{1\over T} \int_{-T/2}^{T/2}\!\! dt \int_{-T/2}^{T/2}\!\!dt'' \, (t-t'') \,  
\lla g{\bf B}({\bf x}_1,t'') \times g{\bf E}({\bf x}_1,t) \rra \nn\\
& & + {c_S^{(1)}\over 2 r^2}{\bf r}\cdot (\bfnabla_r V^{(0)}),
\label{vls20}\\
V_{L_2S_1}^{(1,1)}(r) \!\!&=&\!\! {c_F^{(1)} \over 2r^2}i {\bf r}\cdot \lim_{T\rightarrow \infty}
{1\over T} \int_{-T/2}^{T/2}\!\!dt \int_{-T/2}^{T/2}\!\!dt'' \, (t-t'') \,  
\lla g{\bf B}({\bf x}_1,t'') \times g{\bf E}({\bf x}_2,t) \rra, \label{vls}\\
V_{{\bf p}^2}^{(2,0)}(r) \!\!&=&\!\! {i \over 4}{\hat {\bf r}}^i{\hat {\bf r}}^j
\lim_{T\rightarrow \infty}{1\over T} \int_{-T/2}^{T/2}\!\!dt \int_{-T/2}^{T/2}\!\!dt'' \, (t-t'')^2 \,  
\lla g{\bf E}^i({\bf x}_1,t'') \, g{\bf E}^j({\bf x}_1,t) \rra_c, \label{vp20}\\
V_{{\bf L}^2}^{(2,0)}(r)\!\!&=&\!\! i {\delta^{ij}-3{\hat {\bf r}}^i{\hat {\bf r}}^j \over 8}
\lim_{T\rightarrow \infty}{1\over T} \int_{-T/2}^{T/2}\!\!dt \int_{-T/2}^{T/2}\!\!dt'' \, (t-t'')^2 \,  
\lla g{\bf E}^i({\bf x}_1,t'') g{\bf E}^j({\bf x}_1,t) \rra_c, \label{vl20}\\
V_{{\bf p}^2}^{(1,1)}(r)\!\!&=&\!\!{i\over 2}{\hat {\bf r}}^i{\hat {\bf r}}^j 
\lim_{T\rightarrow \infty}{1\over T} \int_{-T/2}^{T/2}\!\!dt \int_{-T/2}^{T/2}\!\!dt'' \, (t-t'')^2 \,  
\lla g{\bf E}^i({\bf x}_1,t'') \, g{\bf E}^j({\bf x}_2,t) \rra_c, \label{vp11} \\
V_{{\bf L}^2}^{(1,1)}(r) \!\!&=&\!\! i {\delta^{ij}-3{\hat {\bf r}}^i{\hat {\bf r}}^j \over 4}
\lim_{T\rightarrow \infty}{1\over T} \int_{-T/2}^{T/2}\!\!dt \int_{-T/2}^{T/2}\!\!dt'' \, (t-t'')^2 \,  
\lla g{\bf E}^i({\bf x}_1,t'')\, g{\bf E}^j({\bf x}_2,t) \rra_c,\label{vl11}
\eea 
taken from Ref. \cite{m2} (changing, for further convenience,  single in double integrals).
In order to ease the reader we use the same notation as \cite{m2}, 
i.e. the angular brackets $\langle \dots \rangle$ stand for the average value over the
Yang--Mills action, $W_\Box$ for the rectangular static Wilson loop of extension 
$r\times T$ (the time runs from $-T/2$ to $T/2$, the space coordinate from ${\bf x}_1$ 
to ${\bf x}_2$):
\be
W_\Box \equiv {\rm P} \exp\left\{{\displaystyle - i g \oint_{r\times T} \!\!dz^\mu A_{\mu}(z)}\right\},
\qquad 
dz^\mu A_{\mu} \equiv dz^0 A_0 - d{\bf z} \cdot {\bf A},
\ee
which we graphically represent in Fig. \ref{figw}, 
and $\langle\!\langle \dots \rangle\!\rangle 
\equiv \langle \dots W_\Box\rangle / \langle  W_\Box\rangle$; P is the path-ordering operator.
Moreover, we define the connected Wilson loop with $O_1(t_1)$, $O_2(t_2)$ and $O_3(t_3)$ operator 
insertions by: 
\bea
& & \hspace{-5mm} \lla O_1(t_1)O_2(t_2)O_3(t_3)\rra_c = 
\lla O_1(t_1)O_2(t_2)O_3(t_3)\rra - \lla O_1(t_1)\rra \lla O_2(t_2)\rra \lla O_3(t_3)\rra \nn\\
& & \hspace{6mm} - \lla O_1(t_1)\rra \lla O_2(t_2)O_3(t_3)\rra_c 
- \lla O_2(t_2)\rra \lla O_1(t_1)O_3(t_3)\rra_c 
- \lla O_3(t_3)\rra \lla O_1(t_1)O_2(t_2)\rra_c, \nn\\
& & \hspace{-5mm} \lla O_1(t_1)O_2(t_2)\rra_c = \lla O_1(t_1)O_2(t_2)\rra - \lla O_1(t_1)\rra \lla O_2(t_2)\rra. \nn
\eea 
The operators ${\bf E}^i= F_{0i}$ and ${\bf B}^i=\epsilon^{ijk}F^{jk}/2$ 
($F_{\mu\nu} = \partial_\mu A_\nu - \partial_\nu A_\mu + ig[A_\mu,A_\nu]$)
are the chromoelectric and chromomagnetic field respectively.

\begin{figure}[htb]
\makebox[0.5cm]{\phantom b}
\put(110,0){\epsfxsize=5.5truecm \epsfbox{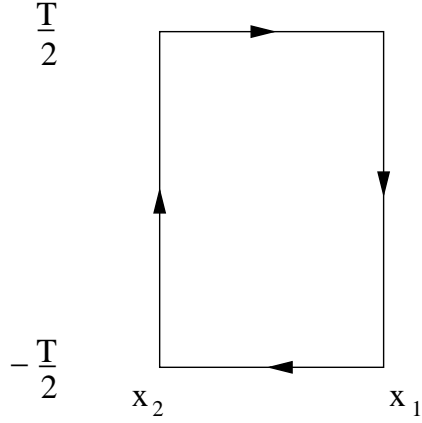}}
\vspace{3mm}
\caption{\it Graphical representation of the static Wilson loop $W_\Box$ of extension $r\times T$.}
\label{figw}
\end{figure}

The matching coefficients $c_F^{(j)}$, $c_S^{(j)}$ are inherited from the matching 
between QCD and NRQCD. Their explicit form does not matter here. The only point relevant 
for our further discussion is that, from reparameterization invariance, one has \cite{manohar}
\be
c_S^{(j)} = 2 c_F^{(j)} -1. 
\label{par}
\ee

\subsection{Relations for spin-dependent potentials}
\label{secsd}
In this section we derive the relations (\ref{spin1})-(\ref{spin2}) 
by explicit transformation of the potentials (\ref{vls20}) and (\ref{vls}). 
The proof is, partially, a simplified version of that one derived long ago by one of the authors \cite{poG,rev}.

We start from the identity 
\begin{equation}  
\lla g{\bf B} ({\bf x}_1,t'')\rra = 0,  
\label{parB}
\end{equation}
which follows from parity.
Then we apply an infinitesimal Lorentz boost with velocity ${\bf v}$, and the point
$({\bf x}_1,t'')$ as the origin in both systems, i.e. up to order ${\cal O}({\rm v})$
\bea
\left\{
\begin{array}{l}
\displaystyle{{\bf x} - {\bf x}_1 = {\bf x}' - {\bf x}_1 + (t'-t''){\bf v}}, \\
\displaystyle{} \\
\displaystyle{t-t'' = t'-t'' + ({\bf x}' - {\bf x}'_1)\cdot{\bf v}}.
\end{array}
\right.
\eea
The Wilson loop transforms like in Fig. \ref{fig2}.
The time-like paths become ${\bf x}'_{1,2}(t') = {\bf x}_{1,2} - (t'-t''){\bf v}$, 
with the boundaries $t_{\pm}^{(1,2)\,\prime} = \pm T/2 - ({\bf x}_{1,2} - {\bf x}_1){\bf v}$. 
To make the last contribution vanish it is convenient to choose ${\bf v} \perp {\bf r}$ in the
following of this section. The spatial lines become 
${\bf x}'_{\pm}(s) = {\bf x}_2 + s({\bf x}_1-{\bf x}_2) - {\bf v}(\pm T/2-t''),\quad 0\leq s \leq 1$. 
In the new system Eq. (\ref{parB}) becomes 
\begin{equation} 
\lla g{\bf B}'({\bf x}_1,t'')\rra^\prime + \lla [{\bf v}\times g{\bf E}'({\bf x}_1,t'')]\rra^\prime =0.
\label{EB0}
\end{equation}
The second term comes from the transformation of the magnetic field,
and the prime on the double brackets denotes that we have to use the transformed Wilson loop.
We next drop the primes for the fields and consider (\ref{EB0}) in the old system. Subtracting Eq. (\ref{parB}) 
from Eq. (\ref{EB0}), dividing by $T$,  and integrating over $t''$ we get at order ${\cal O}({\rm v})$
\begin{equation} 
 \frac{1}{T} \int_{-T/2}^{T/2} dt'' \lla g{\bf B}({\bf x}_1,t'')\rra^\prime 
-\frac{1}{T} \int_{-T/2}^{T/2} dt'' \lla g{\bf B}({\bf x}_1,t'')\rra 
+ \frac{1}{T}\int_{-T/2}^{T/2} dt'' \lla [{\bf v}\times g{\bf E}({\bf x}_1,t'')]\rra = 0.
\label{EB0dif}
\end{equation}
Notice that the second term of (\ref{EB0}) has been just rotated back because it is already 
of order ${\cal O}({\rm v})$. Rewriting the differences in the paths of the first and second term in 
Eq. (\ref{EB0dif}) as insertions of chromoelectric and chromomagnetic fields,
we get, up to order ${\cal O}({\rm v})$ 
\bea 
&& {\bf v}^i\left\{
{i\over T}\int_{-T/2}^{T/2} dt''\int_{-T/2}^{T/2} dt \, (t-t'')
\bigg[ \lla g{\bf B}^k({\bf x}_1,t'')\,g{\bf E}^i({\bf x}_1,t)\rra 
- \lla g{\bf B}^k({\bf x}_1,t'')\,g{\bf E}^i({\bf x}_2,t)\rra \bigg] \right.\nn\\
&&\hspace{7mm} + {i\over T}\epsilon^{ijl}{\bf r}^j\int_0^1 ds \int_{-T/2}^{T/2} dt'' 
\bigg[ \left(t''+ {T\over 2}\right) 
\lla g{\bf B}^k({\bf x}_1,t'')\,g{\bf B}^l({\bf x}_2+s{\bf r},-T/2)\rra 
\nn\\
&&\hspace{50mm}
- \left(t''- {T\over 2}\right) 
\lla g{\bf B}^l({\bf x}_2+s{\bf r},T/2) \, g{\bf B}^k({\bf x}_1,t'')\rra \bigg]\nn\\
&&\hspace{7mm}\left. 
+ {1\over T}\epsilon^{kij} \int_{-T/2}^{T/2} dt'' \lla g{\bf E}^j({\bf x}_1,t'') \rra \right\}=0.
\label{EB0dif2}
\eea

\begin{figure}[\protect{t}h]
\makebox[0.5cm]{\phantom b}
\put(110,0){\epsfxsize=7.5truecm \epsfbox{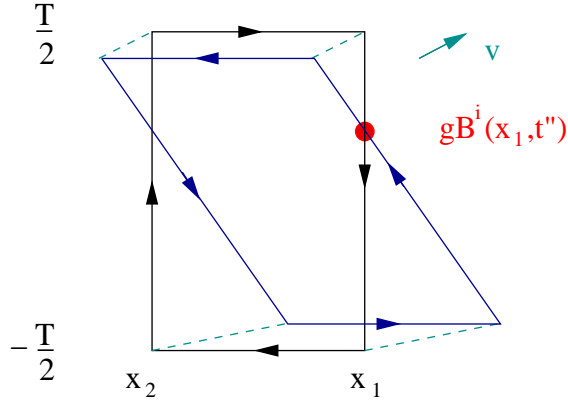}}
\vspace{3mm}
\caption{ \it Difference between the boosted expression (\ref{parB}) and the original one. 
The arrows indicate the orientation of the loops, which arise after expanding up to  first order 
of ${\rm v}$. The corners of the old and the new loop have been connected by dashed lines, in order
to visualize the form of the infinitesimal loops, which correspond to the insertions of chromoelectric 
(time-like triangles at ${\bf x}_1$ and ${\bf x}_2$) and chromomagnetic fields (space-like parallelograms 
at $\pm T/2$).}
\label{fig2}
\end{figure}

Now we consider the large $T$ limit of Eq. (\ref{EB0dif2}). In \cite{m1} it has been proved that 
for $T\to\infty$ 
\be
{1\over T}\int_{-T/2}^{T/2} dt'' \lla g{\bf E}^j({\bf x}_1,t'') \rra = 
{{\bf r}^j \over r}{d\over dr}V^{(0)}(r) + {\cal O}\left({1\over T}\right).
\label{largeT1}
\ee
Moreover, we have for $T\to\infty$  
\bea 
&&\hspace{7mm}  
{i\over T}\int_{-T/2}^{T/2} dt'' \bigg[ \left(t''+ {T\over 2}\right) 
\lla g{\bf B}^k({\bf x}_1,t'')\,g{\bf B}^l({\bf x}_2+s{\bf r},-T/2)\rra 
\nn\\
&&\hspace{20mm}
- \left(t''- {T\over 2}\right) 
\lla g{\bf B}^l({\bf x}_2+s{\bf r},T/2) \,g{\bf B}^k({\bf x}_1,t'')\rra \bigg] = 
\nn\\
&&\hspace{20mm} 
2 \sum_{n\neq0}{a_n {\bf b}_0^l\over a_0^2} {^{(0)}\langle n|g{\bf B}^k({\bf x}_1)|0\rangle^{(0)} 
\over E_n^{(0)} - E_0^{(0)}} + {\cal O}\left({1\over T}\right), 
\label{largeT2}
\eea
where $|n\rangle^{(0)}$ are the eigenstates of the static NRQCD Hamiltonian, $a_n$ is the projection 
on $|n\rangle^{(0)}$ of the state made by two static quarks connected by a straight string, 
and ${\bf b}_n^l$ of the state made by two static quarks and the chromomagnetic 
field $g{\bf B}^l({\bf x}_2+s{\bf r},-T/2)$ connected by a straight string 
(for more specification on this way of analyzing the large $T$ limit of Wilson loops we 
refer to \cite{m1,m2}). Since, due to the different quantum numbers, this latter state has no overlap with 
the ground state $|0\rangle^{(0)}$, which we identify with the color singlet quarkonium 
state $S$ introduced in Sec. \ref{secpnrqcd}, we have ${\bf b}_0^l =0$. Therefore, the left-hand side 
of Eq. (\ref{largeT2}), vanishes in the large $T$ limit.

From the above analysis it follows that Eq. (\ref{EB0dif2}) reduces, in the large $T$ limit, to the identity:
\bea 
&& \lim_{T\rightarrow \infty} 
{i\over T}\int_{-T/2}^{T/2} dt''\int_{-T/2}^{T/2} dt \, (t-t'')
\bigg[ \lla g{\bf B}({\bf x}_1,t'')\times g {\bf E}({\bf x}_1,t)\rra 
- \lla g{\bf B}({\bf x}_1,t'')\times g{\bf E}({\bf x}_2,t)\rra \bigg] \nn\\
&&\hspace{7mm} + 2  {{\bf r} \over r}{d\over dr}V^{(0)}(r) =0.
\label{EB0dif3}
\eea
This corresponds to the relation\footnote{The potentials $V'_1$,  $V'_2$, 
which, up to missing Wilson coefficients not considered in the early literature, 
correspond to the spin-orbit potentials, are defined as 
\begin{eqnarray*}
V'_1(r) &=&  {i\over 2}{{\bf r}\over r} \cdot
\lim_{T\to\infty}{1\over T} \int_{-T/2}^{T/2} dt \int_{-T/2}^{T/2} dt''
(t - t'') \lla g{\bf B} ({\bf x}_1,t'') \times g {\bf E} ({\bf x}_1,t)\rra, \\
V'_2(r) &=&  {i\over 2}{{\bf r}\over r} \cdot
\lim_{T\to\infty}{1\over T} \int_{-T/2}^{T/2} dt \int_{-T/2}^{T/2} dt''
(t - t'') \lla g{\bf B} ({\bf x}_1,t'') \times g {\bf E} ({\bf x}_2,t)\rra.
\end{eqnarray*}
See also appendix \ref{appsub}.}, 
$V'_2  - V'_1 =  V^{(0)\,\prime}$,  given first in Ref. \cite{poG}. 
Moreover, using Eq. (\ref{par}), we obtain Eqs. (\ref{spin1})-(\ref{spin2}).

We would like to remark that this result depends crucially on the fact that 
the contributions from the shifts of the space-like lines of the
Wilson  loop in Eq. (\ref{largeT2}) vanish in the large $T$ limit.
These contributions were apparently never considered in the literature so far and it may 
be regarded as a largely lucky circumstance that they actually do not affect the identities involving 
the spin-dependent potentials. From a more technical point of view, 
this is related to the fact that the Wilson loop operators on the left-hand side of Eq. (\ref{largeT2})
are multiplied by the factors $(t''+ T/2)$ and  $(t''- T/2)$ respectively, which are identical with the
time differences of the correlation functions. In different
situations the corresponding end-point contributions cannot be
neglected. This fact was generally overlooked in the past literature and led to wrong
statements and contradictions. This will be clarified in appendix \ref{appsub}.

\subsection{Relations for spin-independent potentials}
\label{secsi}
In this section we derive the relations (\ref{bmp1})-(\ref{bmp2})
by explicit computation on the potentials (\ref{vp20})-(\ref{vl11}). 
It is convenient, for our purpouses, to introduce the following definitions\footnote{ 
In terms of the potentials defined at the beginning of Sec. \ref{secex}, we have
$$
V_{{\bf p}^2}^{(2,0)} = V_d - {2\over 3}V_e, \qquad 
V_{{\bf L}^2}^{(2,0)} = V_e, \qquad 
V_{{\bf p}^2}^{(1,1)} =-V_b + {2\over 3}V_c, \qquad 
V_{{\bf L}^2}^{(1,1)} =-V_c.
$$} 
\cite{bmp}:
\begin{eqnarray} 
S^{jk} & = & \delta ^{jk}V_b + \left(\frac{\delta^{jk}}{3} - \frac{{\bf r}^j{\bf r}^k}{r^2}\right)V_c 
\nn\\
&=& -\frac{i}{2}\lim_{T\to\infty}{1\over T} 
\int_{-T/2}^{T/2} dt' \int_{-T/2}^{T/2} dt''
(t'-t'')^2 \lla g{\bf E}^j({\bf x}_2,t') \, g{\bf E}^k({\bf x}_1,t'') \rra_c, 
\label{STtensor1}\\
T^{jk} & = & \delta ^{jk}V_d + \left(\frac{\delta ^{jk}}{3} -\frac{{\bf r}^j{\bf r}^k}{r^2}\right)V_e 
\nn\\
&=& \frac{i}{4}\lim_{T\to\infty}{1\over T}  
\int_{-T/2}^{T/2} dt' \int_{-T/2}^{T/2} dt''
(t'-t'')^2 \lla g{\bf E}^j({\bf x}_1,t') \, g{\bf E}^k({\bf x}_1,t'') \rra_c.
\label{STtensor2}
\end{eqnarray}
The above formulae involve insertions of two chromoelectric fields, together with quadratic 
factors in the time differences. Therefore, we have to consider the Lorentz transformation
up to second order in the velocity ${\rm v}$.

We start with an expression like (\ref{parB}), but with a chromoelectric field instead 
of the chromomagnetic one, i.e.
\begin{equation} 
\lla ig{\bf v}\cdot{\bf E}({\bf x}_1,t'') \rra .  
\label{Efield}
\end{equation}
The velocity ${\bf v}$ will be used in the Lorentz boost. 
The longitudinal component of the chromoelectric field which appears in (\ref{Efield})  is
not changed under this transformation, this greatly simplifies the
derivation. Another reason for taking the scalar product with ${\bf v}$ will become obvious below. 
The Lorentz transformation with $({\bf x}_1,t'')$ as origin is now needed up to 
order ${\cal O}({\rm v}^2)$. It reads
\bea
\left\{
\begin{array}{l}
\displaystyle{
{\bf x}-{\bf x}_1 = {\bf x}' - {\bf x}_1 + (t'-t''){\bf v} 
+ \frac{1}{2}[({\bf x}' - {\bf x}_1)\cdot{\bf v}] {\bf v}, } \\
\displaystyle{}    \\
\displaystyle{
t-t'' = \left(1+\frac{v^2}{2}\right)(t'-t'') + ({\bf x}' - {\bf x}_1)\cdot{\bf v}. } 
\end{array}\right.
\eea
In the new system the time-like lines of the Wilson loop become
\begin{equation} 
{\bf x}'_{1,2}(t') = {\bf x}_{1,2} - (t'-t''){\bf v} -
\frac{1}{2}[({\bf x }_{1,2} - {\bf x}_1)\cdot{\bf v}]{\bf v}, 
\label{g44}
\end{equation}
with the boundaries 
\begin{equation} 
t^{(1,2)\,\prime}_\pm = \pm \left(1+\frac{v^2}{2}\right)\frac{T}{2} -
({\bf x}_{1,2} - {\bf x}_1)\cdot{\bf v} - \frac{v^2}{2} t''.
\end{equation}
The spatial lines become 
\be
{\bf x}'_{\pm}(s) = {\bf x}_2 + s({\bf x}_1-{\bf x}_2) - {\bf v}(\pm T/2-t'')
+{1\over 2}\,(1-s)\,[({\bf x}_1-{\bf x}_2)\cdot{\bf v}]{\bf v}, \quad 0\leq s \leq 1. 
\ee
We now write  (\ref{Efield}) in the new coordinate system. Recalling that the longitudinal 
component of ${\bf E}$ is not transformed, we obtain
\begin{equation} 
\lla ig{\bf v} \cdot{\bf E}'({\bf x}_1,t'') \rra'. 
\label{Efield2}
\end{equation}
Again the prime denotes the transformed Wilson loop. 
We next drop the primes for the field in the boosted system and consider the difference between 
the expressions (\ref{Efield2}) and (\ref{Efield}), divide by $T$, and integrate over $t''$ 
from $-T/2$ to $T/2$. This has to vanish by Lorentz invariance: 
\begin{equation} 
\frac{1}{T }\int_{-T/2}^{T/2} dt'' \lla ig{\bf v} \cdot{\bf E}({\bf x}_1,t'') \rra' -
\frac{1}{T} \int_{-T/2}^{T/2} dt'' \lla ig{\bf v} \cdot{\bf E}({\bf x}_1,t'') \rra  = 0. 
\label{EpmE}
\end{equation}
The figure looks like Fig. \ref{fig2} with ${\bf B}({\bf x}_1,t'')$ replaced by ${\bf E}({\bf x}_1,t'')$, 
but the shift of the Wilson loop lines  has now to be considered up to order ${\rm v}^2$. 

\begin{figure}[htb]
\makebox[0.5cm]{\phantom b}
\put(170,0){\epsfxsize=3.5truecm \epsfbox{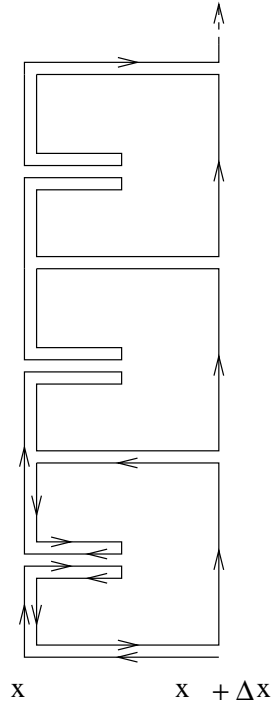}}
\vspace{3mm}
\caption{ \it Distortion of the path showing the appearance of the field insertions and the covariant derivatives.}
\label{fig3}
\end{figure}

In the Abelian case the evaluation of the left-hand side of Eq. (\ref{EpmE}) 
could be easily done according to the self-explaining scheme
\begin{eqnarray} 
\exp \left\{ - ig\int_{C'}\right\} & = &\exp \left\{ - ig\int_C 
- ig \int_{C'-C}\right\} \nonumber\\
& = & \left[\exp \left\{ - ig\int_C\right\}\right] \; \left[1 - ig\int_{C'-C} -
\frac{g^2}{2}\left(\int_{C'-C}\right)^2 + \dots\right], 
\label{stokes}
\end{eqnarray}
where $C$ is the original Wilson loop and $C'$ the boosted one.
More specifically, considering the time-like paths only, we get 
$\int_{C'-C} = \int {\bf E}\,d{\bf a},$ where the integral is now over the area enclosed 
by the time-like lines of the paths $C'$ and $C$. The two paths differ by order ${\rm v}$. 
In order to get the correct result up to order ${\rm v}^2$ one has to expand ${\bf E}({\bf x}_{1,2}
-s(t'-t''){\bf v} - s[{\bf v}\cdot({\bf x}_{1,2} - {\bf x}_1)]{\bf v}/2,t') = 
{\bf E}({\bf x}_{1,2},t') - s(t'-t'')({\bf v} \cdot {\bf \nabla}){\bf E}({\bf x}_{1,2},t') + {\cal O}({\rm v}^2)$. 
Here $s$ parameterizes the transversal direction between the lines $C'$ and $C$; it has to be
integrated from 0 to 1, which gives a factor 1/2 in the second term.
Analogously we may calculate the contributions coming from the shifts in the spatial lines 
(these, however, involve also chromomagnetic fields).

In the non-Abelian case one can apply a method similar to that used in the derivation 
of the non-Abelian Stokes theorem \cite{stokes}. Distort the line at, say, ${\bf x} + \Delta {\bf x}$ 
by adding appendices as shown in Fig. \ref{fig3}. This gives a path at ${\bf x}$, 
together with insertions of chromoelectric fields, represented 
by the paths around the small rectangles. They start and end in the middle of the rectangles and
give the chromoelectric field at the position ${\bf x}+\Delta {\bf x}/2$. 
Together with the small lines connecting the midpoint with the line at ${\bf x}$ one has
\begin{eqnarray} 
e^{-ig{\bf A}({\bf x})\cdot\Delta{\bf x}/2} \,{\bf E}^k\left({\bf x} + {\Delta{\bf x}\over 2}\right) \,
e^{ig{\bf A}({\bf x})\cdot\Delta{\bf x}/2}
&=& {\bf E}^k({\bf x}) + \frac{\Delta {\bf x}}{2} \cdot 
\bigg(({\bfnabla}_{{\bf x}}{\bf E}^k({\bf x})) -
\nonumber\\
ig[{\bf A}({\bf x}),{\bf E}^k({\bf x})]\bigg) + \dots 
&=& {\bf E}^k({\bf x}) + \frac{\Delta {\bf x}}{2}\cdot [{\bf D}_{{\bf x}},{\bf E}^k({\bf x})] + \dots,
\end{eqnarray}
with the covariant derivative ${\bf D}_{{\bf x}}=\bfnabla_{{\bf x}} - i g {\bf A}({\bf x})$.

Applying the above considerations to the left-hand side of Eq. (\ref{EpmE}) one finds that the
contributions of order ${\rm v}^2$ vanish by time reversal invariance, because
the integrals contain factors $(t'-t'')$, which are linear in the time
variables. The spatial shifts do not contribute in the limit $T\rightarrow \infty$. 
This can be seen explicitly by using the same methods and arguments of Sec. \ref{secsd}.
The remaining contributions of order ${\rm v}^3$ are, in the large $T$ limit, 
\begin{eqnarray} 
\hspace{-4mm}
E^{(1,2)} & = & \pm \lim_{T\to\infty} \frac{1}{2T} \int_{-T/2}^{T/2} dt' \int_{-T/2}^{T/2}   dt''
(t'-t'')^2 \nn\\
& & \qquad\qquad\qquad
\times \lla [{\bf v}\cdot{\bf D}_{{\bf x}_{1,2}},g{\bf v}\cdot{\bf E}({\bf x}_{1,2},t')]\, 
g{\bf v}\cdot{\bf E}({\bf x}_1,t'') \rra_c ,\\
\hspace{-4mm}
E^{(3)} & = & - \lim_{T\to\infty} \frac{1}{2T}({\bf r}\cdot{\bf v}) 
\int_{-T/2}^{T/2}  dt' \int_{-T/2}^{T/2}  dt'' 
\lla g{\bf v}\cdot{\bf E}({\bf x}_2,t')\,g{\bf v}\cdot{\bf E}({\bf x}_1,t'') \rra_c,\\
\hspace{-4mm}
E^{(4)} & = & E^{(4)}_{11} +E^{(4)}_{22} - 2 E^{(4)}_{12},  \\
\hspace{-11mm}\mbox{with } ~~~& &\nn \\
\hspace{-4mm}
E^{(4)}_{jk} & = & - \lim_{T\to\infty} \frac{i}{2T} \int_{-T/2}^{T/2}  dt' \int_{-T/2}^{T/2}  dt''  
\int_{-T/2}^{T/2}  dt''' (t'-t'')(t'''-t'') \nn\\
& & \qquad\qquad\qquad \times \lla g{\bf v}\cdot{\bf E}({\bf x}_j,t')\,
g{\bf v}\cdot{\bf E}({\bf x}_k,t''') \,
g{\bf v}\cdot{\bf E}({\bf x}_1,t'') \rra_c .
\end{eqnarray}
The terms $E^{(1,2)}$ arise from the first order shift (with respect to ${\bf v}$) of the paths, 
together with the first order expansion of the chromoelectric field. 
$E^{(3)}$ belongs to the second order shift in (\ref{g44}) at the line ${\bf x}_2$, 
while $E^{(4)}$ is due to the square of the integral in  (\ref{stokes}).

We next rewrite the commutator of the covariant derivative in
$E^{(1,2)}$. This is rather simple for $E^{(2)}$: the loop with the
insertion of $[{\bf D}^j_{{\bf x}_2}, {\bf E}^k({\bf x}_2,t')]$ may be written 
as $\nabla^j_{{\bf x}_2}$ acting on the loop with the insertion of ${\bf E}^k({\bf x}_2,t')$, 
minus $\bfnabla^j_{{\bf x}_2}$ acting on the loop alone. The last
derivative can, in turn, be expressed, in the large $T$ limit,  by an additional insertion of a 
chromoelectric field $ig{\bf E}^j({\bf x}_2,t''')$ and integration over $t'''$.
In this way $E^{(2)}$ becomes a sum of a derivative of a correlator
with two fields and of a correlator with three fields, $E^{(2)} = E^{(2)}_2 + E^{(2)}_3$, with
\begin{eqnarray} 
E^{(2)}_2 & = & - \lim_{T\to\infty} \frac{1}{2T}({\bf v}\cdot{\bfnabla}_{{\bf x}_2}) 
\int_{-T/2}^{T/2} dt' \int_{-T/2}^{T/2} dt'' (t'-t'')^2 \nn\\
& &\qquad\qquad\qquad 
\times \lla g{\bf v}\cdot{\bf E}({\bf x}_2,t')\,g{\bf v}\cdot{\bf E}({\bf x}_1,t'') \rra_c, \label{E10}\\
E^{(2)}_3 & = &  \lim_{T\to\infty} \frac{i}{2T} 
\int_{-T/2}^{T/2}  dt' \int_{-T/2}^{T/2} dt''  \int_{-T/2}^{T/2} dt''' (t'-t'')^2 \nn\\
& &\qquad\qquad\qquad
\times \lla g{\bf v}\cdot{\bf E}({\bf x}_2,t') \,g{\bf v}\cdot{\bf E}({\bf x}_2,t''')\,
g{\bf v}\cdot{\bf E}({\bf x}_1,t'') \rra_c . 
\end{eqnarray}
For $E^{(1)}$ the situation is slightly different,
because $\bfnabla^j_{{\bf x}_1}$ acting on the whole loop together with the
insertions also differentiates ${\bf E}({\bf x}_1,t'')$. The fields
${\bf v}\cdot {\bf E}({\bf x}_1,t')$ and ${\bf v}\cdot {\bf E}({\bf x}_1,t'')$
enter symmetrically, therefore differentiation gives twice the same
expression. Instead of (\ref{E10}) we thus obtain $E^{(1)} = E^{(1)}_2 + E^{(1)}_3$, with
\begin{eqnarray} 
E^{(1)}_2 & = & \lim_{T\to\infty}\frac{1}{4T}({\bf v}\cdot{\bfnabla}_{{\bf x}_1}) 
\int_{-T/2}^{T/2} dt' \int_{-T/2}^{T/2}dt''  (t'-t'')^2 \lla g{\bf v}\cdot{\bf E}({\bf x}_1,t')\,
g{\bf v}\cdot {\bf E}({\bf x}_1,t'') \rra_c, \label{E11}\\
E^{(1)}_3 & = &  \lim_{T\to\infty} \frac{i}{4T} 
\int_{-T/2}^{T/2} dt' \int_{-T/2}^{T/2}dt''  \int_{-T/2}^{T/2}dt''' 
(t'-t'')^2 \nn\\
& & \qquad\qquad\qquad \times  \lla g{\bf v}\cdot{\bf E}({\bf x}_1,t') \,
g{\bf v}\cdot{\bf E}({\bf x}_1,t''')\,g{\bf v}\cdot {\bf E}({\bf x}_1,t'') \rra_c . 
\end{eqnarray}
The correlators $E^{(2)}_2$ and $E^{(1)}_2$ can be directly written in
terms of the expressions $S^{jk}$ and $T^{jk}$ in  (\ref{STtensor1})-(\ref{STtensor2}). 
As intermediate result we obtain 
\begin{equation} 
E^{(2)}_2 + E^{(1)}_2 = i({\bf v}\cdot{\bfnabla}_r) {\bf v}^j (S^{jk} - T^{jk}) {\bf v}^k. \label{E12}
\end{equation}
The contribution $E^{(3)}$ can be expressed by the static
potential $V^{(0)}$ if we rewrite the insertions of the chromoelectric fields as
derivatives acting on the loop. It can be written in the form
\begin{equation} 
E^{(3)} = - \frac{i}{2}({\bf r}\cdot{\bf v})
({\bf v} \cdot {\bfnabla}_r)\frac{({\bf r}\cdot {\bf v})}{r} V^{(0)\,\prime}(r) =
\frac{i}{2} ({\bf v}\cdot{\bfnabla}_r) {\bf v}^j\left\{\delta ^{jk} V^{(0)}(r) -
\frac{{\bf r}^j{\bf r}^k}{r}V^{(0)\, \prime}(r) \right\}{\bf v}^k. 
\label{E13}
\end{equation}
We next consider the expressions which contain insertions with three
fields, namely $E^{(1)}_3,E^{(2)}_3$ and the three terms in $E^{(4)}$. 
In $E^{(4)}_{11}$ we can symmetrize the factor in front with respect to $t'$ and $t''$, 
i.e. replace $(t'-t'')(t'''-t'') \rightarrow (t'-t'')^2/2$. Then one has $E^{(1)}_3 + E^{(4)}_{11} = 0$. 
In $E_3^{(2)}$ and $E^{(4)}_{22}$ we exchange the coordinates ${\bf x}_1$ and ${\bf x}_2$ by applying $CP$,
furthermore we exchange the integration variables $t''$ and $t'''$.
The remaining three terms then all involve the same correlator.
Symmetrising with respect to $t',t''$ we get
\begin{eqnarray} 
\lefteqn{E^{(2)}_3 + E^{(4)}_{22} -2 E^{(4)}_{12} =}
\nonumber\\
& & ~~~\lim_{T\to\infty}\frac{3i}{4T}\int_{-T/2}^{T/2}dt'  \int_{-T/2}^{T/2}dt''  \int_{-T/2}^{T/2}dt'''  \,(t'-t'')^2
\nn\\
& &\qquad\qquad\qquad \times 
\lla g{\bf v}\cdot {\bf E}({\bf x}_1,t') \, g{\bf v}\cdot {\bf E}({\bf x}_1,t'') \,
g{\bf v}\cdot {\bf E}({\bf x}_2,t''') \rra_c \nonumber\\
& & = \lim_{T\to\infty} \frac{3}{4T}({\bf v}\cdot {\bfnabla}_{{\bf x}_2}) 
\int_{-T/2}^{T/2}dt'  \int_{-T/2}^{T/2}dt''\,  (t'-t'')^2
\lla g{\bf v}\cdot {\bf E}({\bf x}_1,t')\,g{\bf v}\cdot {\bf E}({\bf x}_1,t'') \rra_c 
\nonumber\\ & & = 3i({\bf v}\cdot {\bfnabla}_r) {\bf v}^j T^{jk} {\bf v}^k.
\label{E14}
\end{eqnarray}
In the third line we wrote the insertion $\int \cdots ig{\bf v}\cdot{\bf E}({\bf x}_2,t'')\,dt''$ 
as differentiation of the loop with respect to ${\bf x }_2$, in the last line we introduced 
the definition of $T^{jk}$ in  (\ref{STtensor2}). Lorentz invariance requires that the sum of (\ref{E12}),
(\ref{E13}), (\ref{E14}) vanishes. This gives
\begin{equation} 
i({\bf v}\cdot\bfnabla_r) {\bf v}^j\left\{S^{jk} + 2 T^{jk}
+\frac{\delta ^{jk}}{2} V^{(0)}(r) - \frac{{\bf r}^j{\bf r}^k}{2r}V^{(0)\,\prime}(r) \right\}{\bf v}^k = 0.
\label{E15}
\end{equation}
The term in curly brackets is just the expression which has to vanish according to 
Eqs. (\ref{bmp1})-(\ref{bmp2}) and \cite{poBBP}.

Eq. (\ref{E15}) is only slightly weaker than the statement of the vanishing of the curly bracket. 
Let us introduce the combinations
\begin{equation} 
f(r) = V_d(r) + \frac{1}{2}V_b(r) + \frac{1}{4}V^{(0)}(r)
- \frac{r}{12}V^{(0)\,\prime}(r),\quad g(r) = V_e(r) + \frac{1}{2}V_c(r) +
\frac{r}{4}V^{(0)\,\prime}(r).  
\end{equation}
These are the functions that have to vanish according to Eqs. (\ref{bmp1})-(\ref{bmp2}) and 
\cite{poBBP} if $S^{jk}$ and $T^{jk}$ are decomposed into $V_b(r),\cdots ,V_e(r)$. 
In our case (\ref{E15}) gives two differential equations which appear as factors 
of $4i{\rm v}^2({\bf r}\cdot{\bf v})/r^2$ and $4i({\bf r}\cdot{\bf v})^3/r^4$, respectively. They read
\begin{equation} 
rf'(r)/2-g(r) + rg'(r)/6 =0,\quad g(r)-rg'(r)/2 =0.
\end{equation}
The general solution is $g(r)=ar^2$, $f(r)=2ar^2/3+b$ with $a,b=$ constant. Only $a=0$ appears 
physically reasonable, while the constant $b$ may be reabsorbed into the static potential.

\section{Conclusions}
\label{seccon}
In this work we have derived identities among the potentials of pNRQCD 
(as given in Sec. \ref{secpnrqcd}) either by direct application of the Poincar\'e algebra 
to the potentials in their general form or by explicit Lorentz transformation of the potentials 
in terms of Wilson loops. In this way we have proved that 
the identities found long ago for the spin-dependent \cite{poG} and spin-independent 
potentials \cite{poBBP} are correct, despite several subtleties involved 
in the manipulation of Wilson loops overlooked by the literature (as first pointed out 
in \cite{m2} and discussed in appendix \ref{appsub}).
We have thereby also proved that Lorentz invariance is the reason behind these two set 
of identities, providing a unified framework. 

From the general arguments of Sec. \ref{secpo} the identities (\ref{spin1})-(\ref{bmp3}) 
are the only ones that can be derived, between the potentials up to 
order $1/m^2$. The identities that could constrain the potentials of order $1/m$ and the 
momentum independent $1/m^2$ potentials would involve $1/m^3$ potentials, which are, at present, unknown.
In this respect the ``brute force'' method used in Sec. \ref{secex} may be more useful. 
It may be used in order to constrain Wilson loop operators that are relevant for the calculation 
of the potentials up to order $1/m^2$, but do not necessarily show up as potentials themselves. 
In particular, one could apply transformations similar to those discussed in Secs. \ref{secsd} and 
\ref{secsi} on Wilson loops with two or more field insertions. This would result in constraints involving 
Wilson loops with three or more field insertions. These type of Wilson loops are known to contribute 
to the momentum-independent potentials of order $1/m^2$ \cite{m2}.

The complementarity of the two methods discussed in this work, which, however, are both  
applications of Poincar\'e invariance,  can be made clear by noticing that 
the identities (\ref{spin1}) and (\ref{spin2}) proved in Sec. \ref{secpo} 
actually differ from the identity (\ref{EB0dif3}) proved in Sec. (\ref{secsd}).
Indeed, combining Eq. (\ref{EB0dif3}) with Eqs. (\ref{spin1})-(\ref{spin2}) 
we get $c_S^{(j)} = 2 c_F^{(j)} -1$, which, therefore, may be derived in this way 
from Poincar\'e symmetry without using reparameterization invariance arguments.

Finally, we comment on the power-counting issue mentioned in the introduction.
Eqs. (\ref{spin1})-(\ref{bmp2}) contain combinations of $1/m^2$ potentials and the static potential.
Since $V^{(0)}$ scales like $mv^2$ in the heavy quark velocity $v$, as usual in non-relativistic 
bound states, then the combinations of potentials of dimension three appearing in 
Eqs. (\ref{spin1})-(\ref{spin2}) must scale as $m^3v^4$ and those of dimension one 
appearing in Eqs. (\ref{bmp1})-(\ref{bmp2}) as $mv^2$. This result is not trivial.
It implies that the considered potentials are suppressed by an extra power of $v$ 
with respect to the ``natural'' power counting based on their dimension.
In perturbative QCD this extra suppression factor is typically provided by the 
coupling constant, $\als \sim v$. Poincar\'e invariance tells that an extra  suppression factor $v$ 
must be dynamically generated also in the non-perturbative regime.

\bigskip

{\bf Acknowledgements.} 
N.B. and A.V. thank the Institut f\"ur Theoretische Physik of the University of Heidelberg 
for the warm hospitality and the Alexander von Humboldt Foundation for support 
during the first stage of this work.

\vfill\eject

\appendix
\section{Potentials and Lorentz Transformations}
\label{appA}
In this section we show in more detail how the dependence on ${\bf U}^{(0)}$,  
${\bf U}^{(1,0)}$ and ${\bf U}^{(0,1)}$ cancels in the final relations (\ref{spin1})-(\ref{bmp3}).
In order to be definite we will assume 
\bea
{\bf U}^{(0)}({\bf r}) &=& U^{(0)}(r) \, {\bf r},  \\
{\bf U}^{(1,0)}({\bf r},{\bf p}_1) &=&  
\left\{ g(r){{\bf r}\over r}, {\bf p}^2_1\right\} 
+ \{f(r),{\bf p}_1\} + h(r)\,{\bf r}, \\ 
{\bf U}^{(0,1)}({\bf r},{\bf p}_2) &=&  
\left\{ g(r){{\bf r}\over r}, {\bf p}^2_2\right\} 
+ \{f(r),{\bf p}_2\} + h(r)\,{\bf r}, 
\eea
where $g$, $f$ and $h$ are arbitrary functions.
From Eq. (\ref{A9}) it follows that one must have $g(r) = 0$. 
Inserting Eqs. (\ref{UU}), (\ref{K2one}), (\ref{HH}) 
in Eq. (\ref{A7}) we get, up to order $1/m$, the following five equations: 
\begin{eqnarray}
& & 
r {d\over dr}v_b^{(0)}(r) + 2 V_{{\bf L}^2}^{(2,0)}(r) - V_{{\bf L}^2}^{(1,1)}(r)
- r {d \over dr} U^{(0)}(r) =0,
\label{aap1}\\
& & 
r {d\over dr}v_a^{(0)}(r) + 2 V_{{\bf L}^2}^{(0,2)}(r) - V_{{\bf L}^2}^{(1,1)}(r) 
+ r {d \over dr} U^{(0)}(r) =0,
\label{aap4}\\
& & 
-v_a^{(0)}(r) - 2 V_{{\bf L}^2}^{(2,0)}(r) + V_{{\bf L}^2}^{(1,1)}(r)
- 2 V_{{\bf p}^2}^{(2,0)}(r) + V_{{\bf p}^2}^{(1,1)}(r) -  U^{(0)}(r) = 0, 
\label{aap2}\\
& & 
-v_b^{(0)}(r) - 2 V_{{\bf L}^2}^{(0,2)}(r) + V_{{\bf L}^2}^{(1,1)}(r)
- 2 V_{{\bf p}^2}^{(0,2)}(r) + V_{{\bf p}^2}^{(1,1)}(r) +  U^{(0)}(r) = 0, 
\label{aap5}\\
& & 
{f(r)\over r}{d\over dr}V^{(0)}(r) =0.
\label{aap3}
\end{eqnarray}
Adding Eq. (\ref{aap1}) to Eq. (\ref{aap4}) and Eq. (\ref{aap2}) to  Eq. (\ref{aap5})
we get Eqs. (\ref{bmp1}) and (\ref{bmp2}). Subtracting them and combining the result, we get 
Eq. (\ref{bmp3}). Eq. (\ref{aap3})  gives 
\begin{equation}
f(r) =0,
\end{equation}
therefore it follows that ${\bf U}^{(1,0)}$ and ${\bf U}^{(0,1)}$ 
do not depend linearly on ${\bf p}_1$ and ${\bf p}_2$ respectively.
The above derivation may be extended to more general  
functions ${\bf U}^{(1,0)}$ and ${\bf U}^{(0,1)}$ 
that admit a power expansion in the momenta.

\begin{figure}[htb]
\makebox[0.5cm]{\phantom b}
\put(110,0){\epsfxsize=7.3truecm \epsfbox{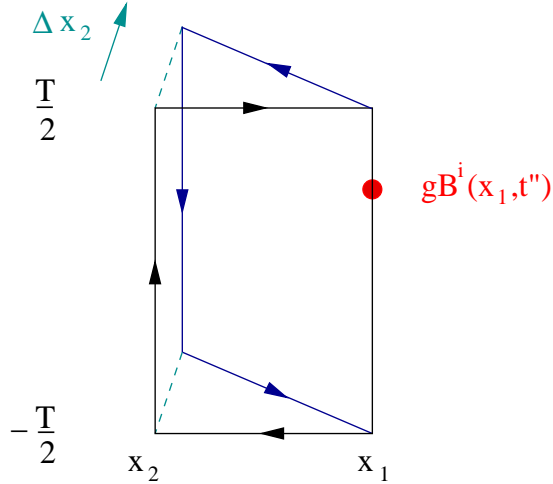}}
\vspace{3mm}
\caption{ \it Difference between the expression (\ref{parB}) shifted accordingly to 
${\bf x}_2 \rightarrow {\bf x}_2 + \Delta {\bf x}_2$ and the original one. 
The arrows indicate the orientation of the loops, which arise after expanding up to first order 
of $\Delta {\bf x}_2$. The corners of the old and the new loop have been connected by dashed lines, 
in order to visualize the form of the infinitesimal loops, which correspond to the insertions of 
chromoelectric (time-like rectangle) and chromomagnetic fields (space-like triangles at $\pm T/2$).}
\label{fig1}
\end{figure}

\section{Subtleties and Incorrect Proofs}
\label{appsub}
In this section we clarify a false statement, due to a wrong treatment of end-point string 
contributions, which can be found in the literature. Let us consider the function \cite{ef,bmp,poG} 
\begin{equation} 
V'_2(r) =  {i\over 2}\epsilon^{kij}{{\bf r}^k\over r} \,
\lim_{T\to\infty}{1\over T} \int_{-T/2}^{T/2} dt \int_{-T/2}^{T/2} dt''
(t - t'') \lla g{\bf B}^i ({\bf x}_1,t'') \, g {\bf E}^j ({\bf x}_2,t)\rra, 
\label{V2p}
\end{equation}
related to the spin-orbit potentials introduced in Sec. \ref{secex} through the relation
$c_F^{(1)} V_2^\prime = r V_{L_2S_1}^{(1,1)}$.
In the review \cite{rev} one finds the statement that the factor $t''$ in
the integrand may be dropped, i.e. that the value of $V_2^\prime$ does not change 
if we replace $(t-t'')$ with $t$. Indeed, in this second form the potential 
$V_2^\prime$ can be found, for instance, in \cite{ef,bmp}.
It has been pointed out and proved in \cite{m2} that this statement is wrong 
even at leading order in perturbation theory. In the following we will make clear how 
it was possible to arrive at that erroneous statement.

Let us start from the identity (\ref{parB}). 
Applying the differential operator $\bfnabla^j_{{\bf x}_2}$ to it and writing the differentiation 
of the time-like lines as insertions of chromoelectric fields and the differentiation of the 
space-like lines as insertions of chromomagnetic fields, we obtain 
\bea
&&  i \int_{-T/2}^{T/2}dt \, \lla g{\bf B}^i({\bf x}_1,t'')\, g{\bf E}^j({\bf x}_2,t)\rra \nn\\
&& -i \epsilon^{jkl}{\bf r}^k \int_{0}^{1}ds \, (1-s) \,  
\lla g{\bf B}^l({\bf x}_2+s{\bf r},T/2)\, g{\bf B}^i({\bf x}_1,t'')\rra \nn\\
&& +i \epsilon^{jkl}{\bf r}^k \int_{0}^{1}ds \, (1-s) \,  
\lla g{\bf B}^i({\bf x}_1,t'')\, g{\bf B}^l({\bf x}_2+s{\bf r},-T/2)\rra =0.
\label{appB1}
\eea
Assuming that the last two lines, once inserted in Eq. (\ref{V2p}), vanish in the large $T$ limit
(as tacitly assumed by previous authors), leads immediately to the conclusion that the factor $t''$ 
may, indeed, be dropped in the integrand of (\ref{V2p}).

The key point is that, as argued in \cite{m2}, the end-point string contributions, corresponding to the 
small triangles at time $\pm T/2$ of Fig. \ref{fig1}, {\it do not vanish} in 
the large $T$ limit in Eq. (\ref{V2p}). This is easy to check at lowest 
order of perturbation theory, where we have:
\bea
&&
 i \int_{-T/2}^{T/2}dt \, \lla g{\bf B}^i({\bf x}_1,t'')\, g{\bf E}^j({\bf x}_2,t)\rra \> 
{\buildrel{=} \over {\hbox{\tiny pert}}} \> 
-2i{C_f\als\over\pi}\epsilon^{ijl}{\bf r}^l \nn\\
&& \hspace{42mm}
\times \left({1\over [(t''-T/2)^2-r^2-i\epsilon]^2} - {1\over [(t''+T/2)^2-r^2-i\epsilon]^2}\right), 
\label{appB2}\\
&&
i \epsilon^{jkl}{\bf r}^k \int_{0}^{1}ds \, (1-s) \,  
\lla g{\bf B}^i({\bf x}_1,t'')\, g{\bf B}^l({\bf x}_2+s{\bf r},- T/2)\rra \>
{\buildrel{=} \over {\hbox{\tiny pert}}} \> 
- 2i{C_f\als\over\pi}\epsilon^{ijl}{\bf r}^l \nn\\
&& \hspace{42mm}
\times {1\over [(t''+ T/2)^2-r^2-i\epsilon]^2},  
\label{appB3}\\
&&
i \epsilon^{jkl}{\bf r}^k \int_{0}^{1}ds \, (1-s) \,  
\lla g{\bf B}^l({\bf x}_2+s{\bf r},T/2)\,g{\bf B}^i({\bf x}_1,t'') \rra \>
{\buildrel{=} \over {\hbox{\tiny pert}}} \> 
- 2i{C_f\als\over\pi}\epsilon^{ijl}{\bf r}^l \nn\\
&& \hspace{42mm}
\times {1\over [(t''-T/2)^2-r^2-i\epsilon]^2},  
\label{appB4}
\eea 
with $C_f = (N_c^2-1)/2 N_c =4/3$ in QCD. It is clear that all three lines  
of Eq. (\ref{appB1}) give the same kind of contribution in leading order perturbation 
theory and that the identity is satisfied only if all three pieces are taken into account.

The argument can be extended to any order of perturbation theory. If we analyze 
the above end-point string contributions by decomposing them into the eigenstates $|n\rangle^{(0)}$
of the static NRQCD Hamiltonian, we obtain for $T \to \infty$ 
\bea 
&&\hspace{-7mm} 
{i\over T}\int_{-T/2}^{T/2} dt'' \, t'' \bigg[ 
\lla g{\bf B}^i({\bf x}_1,t'')\,g{\bf B}^l({\bf x}_2+s{\bf r},-T/2)\rra 
- \lla g{\bf B}^l({\bf x}_2+s{\bf r},T/2) \,g{\bf B}^i({\bf x}_1,t'')\rra \bigg] = 
\nn\\
&&\hspace{20mm} 
{1\over a_0^2}\sum_{n\neq0}{
  a_n {\bf b}_0^l\, {^{(0)}\langle n|g{\bf B}^i({\bf x}_1)|0\rangle^{(0)}}
- a_0 {\bf b}_n^l\, {^{(0)}\langle 0|g{\bf B}^i({\bf x}_1)|n\rangle^{(0)}} 
\over E_n^{(0)} - E_0^{(0)}} + {\cal O}\left({1\over T}\right), 
\label{largeT3}
\eea
with the same notation as in Sec. \ref{secsd}. The ground state $|0\rangle^{(0)}$ has no 
overlap with the state made from two static quarks and a chromomagnetic field 
$g{\bf B}^i({\bf x}_2+s{\bf r},-T/2)$ connected by a straight string, hence ${\bf b}_0^l =0$. 
However, contrary to Eq. (\ref{largeT2}), there are still contributions 
that do not vanish for $T\to\infty$. This general argument definitely 
proves that the end-point string contributions of Eq. (\ref{appB1}) 
do not vanish and, therefore, we cannot drop the factor $t''$ in Eq. (\ref{V2p}).

So, we finally find that a forgotten contribution corrects a wrong
proof where necessary, but cancels in a well-known relation (Eq. (\ref{EB0dif3})), 
where it would be very disturbing. Future manipulations of the above type should, therefore, 
be done with the appropriate care, and old ones revisited appropriately.

\end{document}